\documentclass[12pt]{article}

\begin{document}

\begin{titlepage}

\begin{flushright}

\end{flushright}
\vspace*{1.0cm}
\centerline{\Large\bf Brane Solutions with Tension}
\vspace*{2.4cm}
\centerline{\large Sangheon Yun}
\vspace*{1.8cm}
\centerline{\sl School of Physics}
\vspace*{0.3cm}
\centerline{\sl Korea Institute for Advanced Study}
\vspace*{0.3cm}
\centerline{\sl Hoegiro 87, Dongdaemun-gu, Seoul 130-722, KOREA}
\vspace*{0.3cm}
\centerline{\sl sanhan@kias.re.kr}
\vspace*{2.2cm}
\centerline{\bf ABSTRACT} \vspace*{1cm}
In this note, we apply a special metric ansatz to simplify the
equations of motion for gravitational systems. Then we construct
charged brane solutions in $D=n+p+2$ dimensions which have
spherical symmetry of $S^n$ and translational symmetry along $p$
directions. They are characterized by mass density, uniform
tension and electric/magnetic charges, and nonsingular only for
specific tension. In particular, we find the limits and the
coordinate transformations which reduce the charged brane
solutions to $M2$- and $M5$-branes.
We also obtain the regularity condition for an electrically charged two-brane solution which has two tensions.\\


\end{titlepage}

\section{Introduction}

There is only a very limited family of asymptotically flat,
stationary black hole solutions to Einstein equations in four
dimensions, which has a timelike Killing vector at infinity.
According to no hair theorem, every four-dimensional black hole
formed in a gravitational collapse possesses the properties given
only by its mass, charge and angular momentum, and event horizons
of non-spherical topology are forbidden. In higher dimensions,
however, there exist different kinds of black objects such as
conventional black holes with hyperspherical horizons $S^n$, black
strings/branes, Kaluza-Klein black holes, Kaluza-Klein bubbles and
black tubes. The black branes are solutions which are extended in
extra $p$ spatial dimensions and do not diverge at spatial
infinity
\cite{Duff:1996hp,Horowitz:1991cd,Duff:1993ye,Lu:1995cs,Lu:1995yn,Lu:1995sh}.
Many brane solutions have been constructed for simple truncations
of supergravity theories. A black brane may carry electric or
magnetic charges and couple to dilaton fields just like the black
hole in four dimensions.

Black branes in more than four dimensions are of particular
interest, since they exhibit new behaviors that black holes do not
show. For example, since these brane solutions have translational
invariance, thermodynamics can be extended to hydrodynamics (which
describes long-wavelength deviations from thermal equilibrium).
So, in addition to thermodynamic properties such as temperature
and entropy, black branes possess hydrodynamic characteristics of
continuous fluids : viscosity, diffusion constants, etc. In
\cite{Cavaglia:1997hc,Lee:2006jx}, a neutral, static black string
solution which is characterized by two parameters, mass density
$\lambda$ and tension $\tau$, of the source was obtained. This
type of solutions were extended to black branes in higher
dimensions \cite{Lee:2008zj,Shin:2009}. However, we will show that
those solutions are not singular only for specific values of
tension.

The black branes have complicated metrics, and it is usually
difficult to solve the second-order differential equations of
motion. In this paper, closely following \cite{Shin:2009}, we will
exploit certain metric ansatz to simplify the equations of motion
and to derive the brane solutions with tension. These solutions
can be shown to be characterized by mass density and tension after
some suitable coordinate transformation, and compared with
deformed brane solutions in \cite{Janssen:2007rc}. Since
higher-derivative corrections spoil the simplification in general,
we will not take them into account in this note.

The organization of the paper is as follows. In section 2 we will
gain a brane solution with tension by solving simplified equations
of motion. Then, in section 3, the result is extended to charged
brane cases. And we will consider the limits and the coordinate
transformations which reduce the charged brane solutions to $M2$-
and $M5$-branes in order to show that such solutions might be
considered as singular extensions of $M2$- and $M5$-branes. An
electrically charged two-brane solution with two tensions and its
regularity condition are obtained in section 4. Section 5 is
reserved for the discussions.

\section{Neutral Brane Solutions with Tension}

Let us start with the $(n+p+2)$-dimensional Einstein-Hilbert
action ($n \geq 2$),
\begin{equation}
I=\frac{1}{16\pi G_{\!n+p+2}}\!\int\!d^{n+p+2}\!x\;\sqrt{-g}\;\mathcal{R}.
\end{equation}
We would like to recombine the metric fields. For convenience,
uniformity for all the translationally symmetric directions of the
solution will be assumed. Then we take the metric ansatz
(similarly as in \cite{Janssen:2007rc,Miller:2006ay}) as follows,
\begin{eqnarray}
\label{metricansatz1}
ds^2&=&-e^{\frac{2}{p+1}\{j(r)+p\,k(r)\}}\,dt^2+e^{\frac{2}{p+1}\{j(r)-k(r)\}}\sum_{i=1}^p\,dz_i^2 \nonumber \\
&&~~~+e^{\frac{2}{n-1}\{\,h(r)-j(r)\}}\Big(\,e^{2f(r)}\,dr^2+d\Omega_{n}^{2}\,\Big),
\end{eqnarray}
where $d\Omega_{n}^{2}$ is the metric for $S^{n}$,
$d\Omega_{n}^{2}=d\theta_1^2 +
\sum^{n}_{j=2}\prod^{j-1}_{i=1}\sin^2\theta_i\,d\theta_j^2$. This
gives us the following effective Lagrangian density,
\begin{eqnarray}
\mathcal{L}_\textrm{eff}&=&e^{h(r)-f(r)}\bigg[\frac{(n-1)n}{2}\,e^{2f(r)}+\frac{n}{2(n-1)}\{h^\prime(r)\}^2 -\frac{p}{2(p+1)}\{k^\prime(r)\}^2 \nonumber \\
&&-\frac{n+p}{2(n-1)(p+1)}\{j^\prime(r)\}^2\bigg] - \bigg[\,e^{h(r)-f(r)}\bigg(\frac{n}{n-1}\,h^\prime(r)-\frac{1}{n-1}j^\prime(r)\bigg)\bigg]^\prime.\;\;
\end{eqnarray}
Since there is no derivative term of $f(r)$ in the effective
Lagrangian, it is not a dynamical variable, but simply gives us a
constraint. Thus, we can fix the gauge $f(r)=h(r)$ after varying
the Lagrangian. First varying the effective Lagrangian with
respect to $f(r)$, $h(r)$, $j(r)$ and $k(r)$, we can easily derive
the equations of motion. Then, after choosing the gauge
$f(r)=h(r)$ they are expressed in very simple forms and their
solutions can be readily obtained,
\begin{eqnarray}
\label{metricsol1}
h''(r)=(n-1)^2 e^{2h(r)}\;&\Rightarrow&\;e^{-h(r)}=\frac{n-1}{\kappa}\sinh(\kappa r+c_2), \\
j''(r)=0\;&\Rightarrow&\;j(r)=-\mu r+c_3, \\
k''(r)=0\;&\Rightarrow&\;k(r)=-\nu r+c_4,
\end{eqnarray}
with
\begin{equation}
\label{constrain1}
\frac{n}{n-1}\,\kappa^2 = \frac{n+p}{(n-1)(p+1)}\,\mu^2 + \frac{p}{p+1}\,\nu^2,
\end{equation}
where $\kappa$, $\mu$, $\nu$, $c_2$, $c_3$ and $c_4$ are the
integration constants. Using the definition of $r$ coordinate and
the symmetries of $t$ and $z_i$ rescalings we can set $c_2$, $c_3$
and $c_4$ to vanish. Then, we have a two-parameter family of
solutions. After recombining the two parameters and making the
coordinate transformation in a suitable way given by
\begin{eqnarray}
\rho^{n-1}&=&\bar{M}\coth[(n-1)\bar{M}r]\;,\;\;\bar{M}=\frac{4\pi G_{\!{n+p+2}}M}{\textrm{Vol}(S^n)}\sqrt{\frac{(n-1)(npa^2-2pa+n+p-1)}{n(n+p)}}\,\nonumber \\
\kappa&=&2(n-1)\bar{M}\;,\;\;\mu=-\frac{2b(n-1)^2(pa+1)}{pa-n-p+1}\bar{M},\;\;\nu=\frac{2b(n-1)(n+p)(a-1)}{pa-n-p+1}\bar{M},\nonumber \\
b^2&=&\frac{n(pa-n-p+1)^2}{(n+p)(n-1)(npa^2-2pa+n+p-1)},
\end{eqnarray}
where $a$ is the ratio of mass density to tension, we can identify
the parameters as black brane mass density and tension. The
resultant metric is
\begin{equation}
ds^2=-F(\rho)\,dt^2+G(\rho)\,d\rho^2+\rho^2\,G(\rho)\,d\Omega_n^2+H(\rho)\sum_{i=1}^p\,dz_i^2
\end{equation}
where
\begin{eqnarray}
&&F(\rho)=\Lambda(\rho)^{2b},\;\;G(\rho)=\big(1-\frac{\bar{M}^2}{\rho^{2(n-1)}}\big)^\frac{2}{n-1}\Lambda(\rho)^{\frac{2(pa+1)}{pa-n-p+1}b},\nonumber\\
&&H(\rho)=\Lambda(\rho)^{-\frac{2(na-1)}{pa-n-p+1}b},\;\;\Lambda(\rho)=\frac{\rho^{n-1}-\bar{M}}{\rho^{n-1}+\bar{M}}.
\end{eqnarray}
We can check explicitly that the above solution is identified as
the vacuum $p-$brane with trans-spherical symmetry of
\cite{Lee:2008zj} by equalizing all the $p$ directions of
tensions, and reduces to the boosted black brane in the literature
\cite{Cavaglia:1997hc} by an appropriate coordinate
transformation, $\Lambda(\rho)=e^{-K\bar\Delta x}$, and fixing the
symmetries of $t$ and $z_i$ rescalings and the overall scaling of
the metric,
\begin{eqnarray}
ds^2&\!=\!&-e^{K\delta x}\,dt^2+e^{-2\beta Kx}\sum_{i=1}^p\,dz_i^2 + e^{\frac{nK}{2(n-1)\gamma}\{1+2\gamma\delta(\beta+\frac{1}{4\gamma})\}x} \nonumber \\
&&\cdot\sinh^{-\frac{2n}{n-1}}(K\bar\Delta x)\big[dx^2+\frac{(n-1)^2}{K^2\bar\Delta^2}\sinh^2(K\bar\Delta x)d\Omega_n^2\big],~~~~
\end{eqnarray}
where
\begin{equation}
\bar\Delta^2=\frac{\delta}{4}(\delta-\frac{1}{\gamma})-\frac{1+\beta}{4\gamma}, ~~\beta=-\frac{1}{n+p}, ~~\gamma=-\frac{n(n+p)}{4p}, ~~\delta=-\frac{2(pa-n-p+1)}{(n+p)(na-1)}.
\end{equation}

Note that the above brane solution covers only exterior region of
the outer horizon, which is located at $r=\infty$. To avoid a
conical singularity at the event horizon imposes another condition
on the parameters,
\begin{equation}
\label{constrai1}
n(p+1)\kappa = (n+p)\mu + p(n-1)\nu.
\end{equation}
Then (\ref{constrain1}) and (\ref{constrai1}) imply $\mu=\nu$.
Thus, the tension parameter must have a definite value for the
solution to be regular on the horizon.

\section{Dyonic Solutions with Tension}

The extension of the previous result to the charged solutions can
be easily done. The result may be compared with the known metric
for M-branes in eleven dimensions. Let us consider the
$(n+p+2)$-dimensional Einstein-Hilbert-Maxwell action ($n \geq
2$),
\begin{equation}
S=\int \frac{d^{n+p+2}x \sqrt{-g}}{16\pi G_{n+p+2}}\bigg[ \mathcal{R}-\frac{F^{A\mu_1\cdots\mu_{p+2}}F^A_{~\mu_1\cdots\mu_{p+2}}}{2(p+2)!} - \frac{\mathcal{F}^{B\nu_1\cdots\nu_{n}}\mathcal{F}^B_{~\nu_1\cdots\nu_{n}}}{2n!} \bigg],
\end{equation}
where $A$ runs over $A=1,\cdots,N_e$ and $B$ over
$B=1,\cdots,N_m$. $N_e$ is the number of electric fields and $N_m$
is that of magnetic ones. As in the previous case, we recombine
the metric fields and take the same ansatz as
(\ref{metricansatz1}). We want to solve the Maxwell equations
first, and the Einstein equations next. The Maxwell equations are
given by
\begin{eqnarray}
\partial_{\mu_1} \left[ \sqrt{-g} F^{A\,\mu_1\cdots\mu_{p+2}} \right] =  0,\;\;\;
\partial_{\nu_1} \left[ \sqrt{-g} \mathcal{F}^{B\,\nu_1\cdots\nu_{n}} \right] =  0,
\end{eqnarray}
and they can be easily solved,
\begin{eqnarray}
\label{fieldstrength2}
F^A_{~trz_1\cdots z_{p}}&=&Q^A\;e^{f(r)-h(r)+2j(r)},\;\;\; \nonumber \\
\mathcal{F}^B_{~\theta_1\cdots\theta_{n}}&=&P^B \;\mathbf{\omega}_{n},
\end{eqnarray}
where $Q^A$ and $P^B$ are electric and magnetic charges,
respectively and $\mathbf{\omega}_{n}$ stands for the volume form
of $S^n$. The Einstein equations are
\begin{eqnarray}
\label{einstein2}
&&R_{\mu\nu}-\frac{1}{2}\mathcal{R}g_{\mu\nu} + \frac{1}{4}g_{\mu\nu}\bigg(\frac{1}{(p+2)!}F^{A}_{~\mu_1\cdots\mu_{p+2}}F^{A\,\mu_1\cdots\mu_{p+2}} + \frac{1}{n!}\mathcal{F}^{B}_{\nu_1\cdots\nu_{n}}\mathcal{F}^{B\, \nu_1\cdots\nu_{n}}\bigg) \nonumber \\
&& - \frac{1}{2(p+1)!}F^{A}_{\;\;\;\mu\mu_2\cdots\mu_{p+2}}F^{A\;\;\mu_2\cdots\mu_{p+2}}_{\;\;\;\nu} - \frac{1}{2(n-1)!}\mathcal{F}^{B}_{\;\;\;\mu\nu_2\cdots\nu_{n}}\mathcal{F}^{B\;\;\nu_2\cdots\nu_{n}}_{\;\;\;\nu}=0.
\end{eqnarray}
Since the action does not have a $f^\prime(r)$ term, it is easily
found that after plugging the results (\ref{fieldstrength2}) into
the Einstein equations (\ref{einstein2}) and appropriately
recombining them one equation is merely a constraint which is
given by
\begin{eqnarray}
\label{constraints2}
&&(n-1)n e^{2f(r)}-\frac{n}{n-1}\{h^\prime(r)\}^2+\frac{n+p}{(n-1)(p+1)}\{j^\prime(r)\}^2 \nonumber\\
&&-\frac{(n-1)(n+p)}{p+1}q^2 e^{2j(r)}+\frac{p}{p+1}\{k^\prime(r)\}^2=0
\end{eqnarray}
with
\begin{eqnarray}
q=\sqrt{\frac{p+1}{2(n-1)(n+p)}\bigg[\sum_{A=1}^{N_e}(Q^A)^2+\sum_{B=1}^{N_m}(P^B)^2\bigg]}.
\end{eqnarray}
Then fixing the gauge $f(r)=h(r)$, the remaining recombined
Einstein equations are given as follows,
\begin{equation}
h''(r)=(n-1)^2 e^{2h(r)},~~j''(r)=(n-1)^2q^2 e^{2j(r)}~~~\textrm{and}~~~k''(r)=0.
\end{equation}
As in the previous section all the Einstein equations could have
been alternatively obtained from variation of the effective
Lagrangian,
\begin{eqnarray}
\mathcal{L}_\textrm{eff}&=&e^{h(r)-f(r)}\bigg[\frac{(n-1)n}{2}\,e^{2f(r)}+\frac{n}{2(n-1)}\{h^\prime(r)\}^2 -\frac{p}{2(p+1)}\{k^\prime(r)\}^2 \nonumber \\
&&-\frac{n+p}{2(n-1)(p+1)}\{j^\prime(r)\}^2-\frac{(n+p)(n-1)}{2(p+1)}q^2e^{2\{f(r)-h(r)+j(r)\}}\bigg] \nonumber \\
&&-\bigg[\,e^{h(r)-f(r)}\bigg(\frac{n}{n-1}\,h^\prime(r)-\frac{1}{n-1}j^\prime(r)\bigg)\bigg]^\prime,
\end{eqnarray}
where the sign of the $Q^2$-term is flipped \cite{Janssen:2007rc}.
The solutions of the equations of motion are
\begin{eqnarray}
\{h^\prime(r)\}^2=(n-1)^2 e^{2h(r)}+\kappa^2\;&\Rightarrow&\;e^{-h(r)}=\frac{n-1}{\kappa}\sinh(\kappa r+c_2), \\
\{j^\prime(r)\}^2=(n-1)^2 q^2 e^{2j(r)}+\mu^2\;&\Rightarrow&\;e^{-j(r)}=\frac{(n-1)q}{\mu}\sinh(\mu r+c_3),\;\; \\
k^\prime(r)+\nu=0\;&\Rightarrow&\;k(r)=-\nu r+c_4,
\end{eqnarray}
where $\kappa$, $\mu$, $\nu$, $c_2$, $c_3$ and $c_4$ are the
integration constants and we used the fact that $\kappa^2$ is
positive,
$\frac{n}{n-1}\kappa^2=\frac{n+p}{(n-1)(p+1)}\mu^2+\frac{p}{p+1}\nu^2$
from the constraint (\ref{constraints2}). Here $c_2$ and $c_4$ can
be set to zero, and $c_3$ to
$\textrm{arcsinh}\Big(\frac{\mu}{(n-1)q}\Big)$ by the definition
of $r$ coordinate and the symmetries of $t$ and $z_i$ coordinate
rescalings. Then, we are left with a $(2+N_e+N_m)$-parameter
family of solutions. More generally, unless we assume the
uniformity, a solution with $p$ nonuniform tensions has
$1+p+N_e+N_m$ parameters.

Now let us take the electrically charged brane with $p=2$ and
$n=7$ into account. This solution may be considered as a singular
extension of the $M2$-brane in that setting $\mu=\nu$, we can find
a coordinate transformation from our electrically charged brane
with $p=2$ and $n=7$ to the nonextremal $M2$-brane as follows,
\begin{equation}
e^{-2\kappa r} = 1-\frac{\kappa}{3\rho^6}
\end{equation}
where $\kappa=\mu=\nu$ and $\rho$ is the radial coordinate.
Similarly, we would like to consider the magnetically charged
brane with $p=5$ and $n=4$. This case could be considered as a
singular generalization of the $M5$-brane in that for the case
$\mu=\nu$, there is a coordinate transformation from the
magnetically charged brane with $p=5$ and $n=4$ to the nonextremal
$M5$-brane as follows,
\begin{equation}
e^{-2\kappa r} = 1-\frac{2\kappa}{3\rho^3}.
\end{equation}
More generically, $M2$($M5$)-brane solutions can be extended to
have two(five) tension parameters along the extended spatial
directions as in \cite{Lee:2008zj} in the same method.

\section{Charged Two-Brane Solution with Two Tensions}

In this section we will repeat the previous calculations in a
similar way to derive an electrically charged two-brane solution
with two tensions. Consider the following $(n+4)$-dimensional
Einstein-Hilbert-Maxwell action ($n \geq 2$),
\begin{equation}
I=\frac{1}{16\pi G_{n+4}}\int d^{n+4}x\;\sqrt{-g}\;\bigg[\mathcal{R}-\frac{F^{\alpha\beta\gamma\delta}F_{\alpha\beta\gamma\delta}}{48}\bigg].
\end{equation}
On the contrary to the previous cases, we will not assume
uniformity for all the translationally symmetric directions of the
solution. Then we take the similar, but slightly different metric
ansatz,
\begin{eqnarray}
\label{metricansatz2}
ds^2&=&-e^{\frac{2}{3}\{j(r)+2\,k(r)\}}\,dt^2+e^{\frac{2}{3}\{j(r)-k(r)\}}\,\bigg(e^{l(r)}dz_1^2+e^{-l(r)}dz_2^2\bigg) \nonumber \\
&&~~~+e^{\frac{2}{n-1}\{\,h(r)-j(r)\}}\Big(\,e^{2f(r)}\,dr^2+d\Omega_{n}^{2}\,\Big).
\end{eqnarray}
We can solve the Maxwell and the Einstein equations in a similar
way to the previous section. Alternatively, we may first plug the
solution of the Maxwell equations into the action with flipping
the sign of the Maxwell term to find the following effective
Lagrangian density,
\begin{eqnarray}
\mathcal{L}_\textrm{eff}&=&e^{h(r)-f(r)}\bigg[\frac{(n-1)n}{2}\,e^{2f(r)} + \frac{n}{2(n-1)}\{h^\prime(r)\}^2 - \frac{1}{3}\{k^\prime(r)\}^2 \nonumber \\
&&- \frac{n+2}{6(n-1)}\{j^\prime(r)\}^2 - \frac{1}{4}\{l^\prime(r)\}^2 -\frac{(n+2)(n-1)}{6}q^2e^{2\{f(r)-h(r)+j(r)\}} \bigg] \nonumber \\
&&- \bigg[\,e^{h(r)-f(r)}\bigg(\frac{n}{n-1}\,h^\prime(r) - \frac{1}{n-1}j^\prime(r)\bigg)\bigg]^\prime,
\end{eqnarray}
where $q=\sqrt{\frac{3}{2(n-1)(n+2)}Q^2}$ and $Q$ is the electric
charge, and then solve the equations derived from the variation.
The solutions of the equations of motion with the gauge
$f(r)=h(r)$ are
\begin{eqnarray}
e^{-h(r)}&=&\frac{n-1}{\kappa}\sinh(\kappa r+c_2), \\
e^{-j(r)}&=&\frac{(n-1)q}{\mu}\sinh(\mu r+c_3), \\
k(r)&=&-\nu r+c_4, \\
l(r)&=&-\sigma r+c_5
\end{eqnarray}
with the constraint,
\begin{equation}
\label{constrain2}
\frac{n}{n-1}\,\kappa^2-\frac{n+2}{3(n-1)}\,\mu^2-\frac{2}{3}\,\nu^2-\frac{1}{2}\,\sigma^2=0,
\end{equation}
where $\kappa$, $\mu$, $\nu$, $\sigma$, $c_2$, $c_3$, $c_4$ and
$c_5$ are the integration constants. Using the definition of $r$
coordinate and the symmetries of $t$, $z_1$ and $z_2$ rescalings
we can set $c_2$, $c_4$ and $c_5$ to vanish, and $c_3$ to
$\textrm{arcsinh}\Big(\frac{\mu}{(n-1)q}\Big)$. Then, we get a
four-parameter family of solutions. These four parameters are some
combinations of mass density, charge and two tensions of the
solution. It is straightforward to find generalized $M5$-brane
solutions with five different tensions by the similar process.

For the above solution, the event horizon is located at
$r=\infty$. The regularity condition requires
\begin{equation}
\label{constrai2}
3n\kappa = (n+2)\mu + 2(n-1)\nu.
\end{equation}
Then (\ref{constrain2}) and (\ref{constrai2}) give us $\mu=\nu$
and $\sigma=0$. Thus, for the solution to have a regular horizon
the tension parameters cannot be arbitrary.

\section{Discussions}

In this paper, we have obtained dyonic brane solutions with
tension. The parameters are mass density and tension, and
electric/magnetic charges. We also have gotten an electrically
charged two-brane solution with two arbitrary tensions, which,
particularly, could be considered as a singular extension of the
nonextremal $M2$-brane. It is straightforward to generalize the
charged brane solutions further to have $p$ different tensions
along $p$ directions with translational symmetry and to find a
generalized $M5$-brane solution with five different tensions. The
brane solutions with uniform tension are the deformed ones which
were already found in \cite{Janssen:2007rc}, where the first-order
formalism \cite{Miller:2006ay} was employed. It would be
interesting to derive solutions which have $p$ different tensions
along $p$ extended directions. More importantly, however, for most
values of tensions, these solutions represent just naked
singularities \cite{Lee:2008zj,Kim:2007ek,Cho:2007md,Gwak:2008sg}.
In other words, the regularity conditions narrow meaningful
solutions and require definite values of tension parameters.
Therefore, it would be interesting to extend the charged brane
solution to have $p$ different arbitrary tensions along $p$
translationally symmetric directions and to seek the role of the
singular black brane solutions. We leave it for future study.

\section*{Acknowledgments}

The author acknowledges stimulating discussions with H.~Yavartanoo
and K.~K.~Kim.


\begin{thebibliography}{0}

\bibitem{Duff:1996hp}
  M.~J.~Duff, H.~Lu and C.~N.~Pope,
  ``The black branes of M-theory,''
  Phys.\ Lett.\  B {\bf 382}, 73 (1996)
  [arXiv:hep-th/9604052].

\bibitem{Horowitz:1991cd}
  G.~T.~Horowitz and A.~Strominger,
  ``Black strings and P-branes,''
  Nucl.\ Phys.\  B {\bf 360}, 197 (1991).

\bibitem{Duff:1993ye}
  M.~J.~Duff and J.~X.~Lu,
  ``Black and super p-branes in diverse dimensions,''
  Nucl.\ Phys.\  B {\bf 416}, 301 (1994)
  [arXiv:hep-th/9306052].

\bibitem{Lu:1995cs}
  H.~Lu, C.~N.~Pope, E.~Sezgin and K.~S.~Stelle,
  ``Stainless super p-branes,''
  Nucl.\ Phys.\  B {\bf 456}, 669 (1995)
  [arXiv:hep-th/9508042].

\bibitem{Lu:1995yn}
  H.~Lu and C.~N.~Pope,
  ``p-brane Solitons in Maximal Supergravities,''
  Nucl.\ Phys.\  B {\bf 465}, 127 (1996)
  [arXiv:hep-th/9512012].

\bibitem{Lu:1995sh}
  H.~Lu and C.~N.~Pope,
  ``Multi-scalar p-brane solitons,''
  Int.\ J.\ Mod.\ Phys.\  A {\bf 12}, 437 (1997)
  [arXiv:hep-th/9512153].

\bibitem{Cavaglia:1997hc}
  M.~Cavaglia,
  ``Two-dimensional reduced theory and general static solution for uncharged black p-branes,''
  Phys.\ Lett.\  B {\bf 413}, 287 (1997)
  [arXiv:hep-th/9709055].

\bibitem{Lee:2006jx}
  C.~H.~Lee,
  ``Black string solutions with arbitrary tension,''
  Phys.\ Rev.\  D {\bf 74}, 104016 (2006)
  [arXiv:hep-th/0608167].

\bibitem{Lee:2008zj}
  J.~Lee, G.~Kang and H.~C.~Kim,
  ``Geometrical properties of the trans-spherical solutions in higher dimensions,''
  arXiv:0801.0482 [gr-qc].

\bibitem{Shin:2009}
  I.~Shin and S.~Yun,
  ``Einstein-Maxwell-Dilaton Black Branes with Arbitrary Tension and Dilaton Charge,''
  Mod. Phys. Lett. A, Vol. 24, No. 10 (2009) pp. 713-720

\bibitem{Janssen:2007rc}
  B.~Janssen, P.~Smyth, T.~Van Riet and B.~Vercnocke,
  ``A first-order formalism for timelike and spacelike brane solutions,''
  JHEP {\bf 0804}, 007 (2008)
  [arXiv:0712.2808 [hep-th]].

\bibitem{Miller:2006ay}
  C.~M.~Miller, K.~Schalm and E.~J.~Weinberg,
  ``Nonextremal black holes are BPS,''
  Phys.\ Rev.\  D {\bf 76}, 044001 (2007)
  [arXiv:hep-th/0612308].

\bibitem{Kim:2007ek}
  H.~C.~Kim and J.~Lee,
  ``Extraordinary vacuum black string solutions,''
  Phys.\ Rev.\  D {\bf 77}, 024012 (2008)
  [arXiv:0708.2469 [gr-qc]].

\bibitem{Cho:2007md}
  I.~Cho, G.~Kang, S.~P.~Kim and C.~H.~Lee,
  ``Spacetime structure of 5D hypercylindrical vacuum solutions with tension,''
  J.\ Korean Phys.\ Soc.\  {\bf 53}, 1089 (2008)
  [arXiv:0709.1021 [gr-qc]].

\bibitem{Gwak:2008sg}
  B.~Gwak, B.~H.~Lee and W.~Lee,
  ``Geodesic Properties and Orbits in 5-dimensional Hypercylindrical Spacetime,''
  arXiv:0806.4320 [gr-qc].


\end{thebibliography}
\end{document}